\documentclass[preprint,superscriptaddress]{revtex4}
\usepackage{graphicx,array,amsfonts,times,verbatim,subfigure,color}
\usepackage{amssymb,amsmath,multirow,rotate,mathrsfs,booktabs}
\usepackage{multirow,epsfig,threeparttable,chngpage,latexsym,dcolumn}
\usepackage{graphics,graphicx,bm,fleqn,epic,eepic,float}
\definecolor{red}{rgb}{1,0,0}

\definecolor{blue}{rgb}{0,0,1}

\begin{document}

\title{Controlling herding in minority game systems}

\author{Ji-Qiang Zhang}
\affiliation{Institute of Computational Physics and Complex Systems, Lanzhou University, Lanzhou Gansu 730000, China}

\author{Zi-Gang Huang} 
\affiliation{Institute of Computational Physics and Complex Systems, Lanzhou University, Lanzhou Gansu 730000, China}
\affiliation{School of Electrical, Computer, and Energy Engineering, Arizona State University, Tempe, AZ 85287, USA}

\author{Zhi-Xi Wu}
\affiliation{Institute of Computational Physics and Complex Systems, Lanzhou University, Lanzhou Gansu 730000, China}

\author{Riqi Su}
\affiliation{School of Electrical, Computer, and Energy Engineering, Arizona State University, Tempe, AZ 85287, USA}

\author{Ying-Cheng Lai}
\affiliation{School of Electrical, Computer, and Energy Engineering, Arizona State University, Tempe, AZ 85287, USA}
\affiliation{Department of Physics, Arizona State University, Tempe, AZ 85287, USA}

\date\today

\begin{abstract}

{\bf Resource allocation takes place in various types of real-world complex
systems such as urban traffic, social services institutions, economical
and ecosystems. Mathematically, the dynamical process of complex resource
allocation can be modeled as {\em minority games} in which the number of
resources is limited and agents tend to choose the less used resource
based on available information. Spontaneous evolution of the resource
allocation dynamics, however, often leads to a harmful herding behavior
accompanied by strong fluctuations in which a large majority of agents
crowd temporarily for a few resources, leaving many others unused.
Developing effective control strategies to suppress and eliminate herding
is an important but open problem. Here we develop a pinning control method.
That the fluctuations of the system consist of intrinsic and systematic
components allows us to design a control scheme with separated control
variables. A striking finding is the universal existence of an optimal
pinning fraction to minimize the variance of the system, regardless of
the pinning patterns and the network topology. We carry out a detailed
theoretical analysis to understand the emergence of optimal pinning and
to predict the dependence of the optimal pinning fraction on the
network topology. Our theory is generally applicable to systems with
heterogeneous resource capacities as well as varying control
and network topological parameters such as the average degree
and the degree distribution exponent. Our work represents a general
framework to deal with the broader problem of controlling collective
dynamics in complex systems with potential applications in social,
economical and political systems.}

\end{abstract}

\maketitle

Resource allocation is an essential process in many real-world systems
such as ecosystems of various sizes, transportation systems (e.g.,
Internet, urban traffic grids, rail and flight networks), public
service providers (e.g., marts, hospitals, and schools), and social
and economic organizations (e.g., banks and financial markets). The
underlying system that supports resource allocation often contains
a large number of interacting components or agents on a hierarchy of
scales, and there are multiple resources available for each agent. As
a result, complex behaviors are expected to emerge ubiquitously in
the dynamical evolution of resource allocation. In particular, in a typical
situation, agents or individuals possess similar capabilities in information
processing and decision making, and they share the common goal of
pursuing as high payoffs as possible. The interactions among the
agents and their desire to maximize payoffs in competing for limited
resources can lead to vast complexity in the system dynamics.

Given resource-allocation system that exhibits complex dynamics, a
defining virtue of optimal performance is that the available resources
are exploited evenly or uniformly by all agents in the system. In contrast,
an undesired or even catastrophic behavior is the emergence
of herding, in which a vast majority of agents concentrate on a few
resources, leaving many other resources idle or unused~\cite{PBC:2000,
Vazquez:2000,GL:2002,ZWZYL:2005,EZ:2000,LK:2004,WYZJZW:2005,ZYZXLW:2005,
HWGW:2006,HZDHL:2012,ZHDHL:2013,DHHL:2014}. If this behavior is not
controlled, the few focused resources would be depleted, possibly
directing agents to a different but still small set of resources.
From a systems point of view, this can lead to a cascading type of
failures as resources are being depleted one after another, eventually
resulting in a catastrophic breakdown of the system on a global
scale. In this paper, we analyze and test an effective strategy to
control herding dynamics in complex resource-allocation systems.

A universal paradigm to model and understand the interactions
and dynamical evolutions in many real world systems is complex
adaptive systems~\cite{Kauffman:book,Levin:1998,AAP:book}, among
which minority game (MG)~\cite{CZ:1997,CMZ:2013} stands out as
a particularly pertinent framework for resource allocation.
MG dynamics was introduced by Challet and Zhang to address the
classic El Farol bar-attendance problem conceived by
Arthur~\cite{Arthur:1994}. In an MG system, each agent makes choice
(e.g., $+$ or $-$, to attend a bar or to stay at home) based on
available global information from the previous round of interaction.
The agents who pick the minority strategy are rewarded, and those
belonging to the majority group are punished due to limited resources.
The MG dynamics has been studied extensively in the past~\cite{CM:1999,
CMZ:2000,MMM:2004,BMM:2007,RMR:1999,KSB:2000,Slanina:2000,ATBK:2004,
JHH:1999,HJJH:2001,LCHJ:2004,TCHJ:2005,CMM:2008,BMFM:2008,XWHZ:2005a,
ZZZH:2005,MR:2002,CMZ:2013,CHWL:2015}.

To analyze, understand, and exploit the MG dynamics, there are two
theoretical approaches: mean field approximation and Boolean
dynamics. The mean field approach was mainly developed by
researchers from the statistical-physics community to cast the MG
problem in the general framework of non-equilibrium phase
transitions~\cite{CZ:1997,Moro:2004,YZ:2008,CMZ:2013}. In the
Boolean dynamics, for any agent, detailed information
about the other agents that it interacts with is assumed to be
available, and the agent responds accordingly~\cite{PBC:2000,
Vazquez:2000,GL:2002,ZWZYL:2005, HWGW:2006,HZDHL:2012,ZHDHL:2013,
DHHL:2014}. Both approaches can lead to ``better than random''
performance in resource utilization. However, herding behavior
in which many agents take identical action~\cite{DIMCCWK:2008} can
also take place, which has been extensively studied and recognized
as one important factor contributing to the origin of complexity
that leads to enhanced fluctuations and, consequently, to
significant degradation in efficiency~\cite{PBC:2000,Vazquez:2000,
GL:2002,ZWZYL:2005,EZ:2000,LK:2004,WYZJZW:2005,ZYZXLW:2005,HWGW:2006,
HZDHL:2012,ZHDHL:2013,DHHL:2014}.

The control strategy we analyze in this paper is the pinning method
that has been studied in controlling the collective dynamics, such
as synchronization, in complex networks~\cite{WC:2002,LWC:2004,CLL:2007,
XLCCY:2007,TWF:2009,PF:2009,YCL:2009,ZHDHL:2013}. For the general
setting of pinning control, the two key parameters are the ``pinning
fraction,'' the fraction of agents chosen to hold a fixed strategy, and
the ``pinning pattern,'' the configuration of plus or minus
strategies assigned to the pinned agents. Our previous
work~\cite{ZHDHL:2013} treated the special case of two resources
of identical capacities, where the pinning pattern was such that
the probabilities of agents pinned to positive or negative strategies
(to be defined later) are equal. Here, we investigate a more realistic
model setting and articulate a general mathematic control framework.
A striking finding is that biased pinning control pattern can lead to
an optimal pinning fraction for a variety of network topologies, so
that the system efficiency can be improved remarkably. We develop a
theoretical analysis based on the mean-field approximation to understand the
non-monotonic behavior of the system efficiency about the optimal pinning
fraction. We also study the dependence of the optimal fraction on the
topological features of the system, such as the average degree and
heterogeneity, and obtain a theoretical upper bound of the system
efficiency. The theoretical predictions are validated with extensive
numerical simulations. Our work represents a general framework
to optimally control the collective dynamics in complex MG systems with
potential applications in social, economical and political systems. \\
\\ \noindent
{\large\textbf{Results}}
\vspace*{0.1in}
\paragraph*{\bf Boolean dynamics.}
In the original Boolean system, a population of $N$ agents compete for
two alternative resources, denoted as $r=+$ and $r=-$, which have
the same accommodating capacity $C_{+}=C_{-}=N/2$. Similar to the
MG dynamics, only the agents belonging to the \emph{global minority}
group are rewarded by one unit of payoff. As a result, the profit of
the system is equal to the number of agents selecting the resource
with attendance less than the accommodating capacity, which constitute
the global-minority group. The dynamical variable of the Boolean system is
denoted as $A_t$, the number of $+$ agents in the system at time step
$t$. The variance of $A_t$ about the capacity $C_{+}$ characterizes
the efficiency of the system. The densities of the $+$ and $-$ agents
in the whole system are $\rho_{+}=A_{t}/N$ and $\rho_{-}=1-\rho_{+}$,
respectively. The state of the system can be conveniently specified by
the column vector $\hat{\rho}=(\rho_{+},\rho_{-})^{\mathrm{T}}$.

A Boolean system has two states (a binary state system), in which agents make
decision according to the local information from immediate neighbors. The
neighborhood of an agent is determined by the connecting structure of
the underlying network. Each agent receives inputs from its neighboring
agents and updates its state according to the Boolean function, a function
that generates either $+$ and $-$ from the inputs~\cite{GL:2002}.
Realistically, for any agent, global information about the minority choice
from all other agents at the preceding time step may not be available.
Under this circumstance, the agent attempts to decide the global
minority choice based on neighbors' previous strategies. To be concrete,
we assume~\cite{ZHDHL:2013,ZWZYL:2005} that agent $i$ with $k_i$
neighbors chooses $+$ at time step $t+1$ with the probability
\begin{equation} \label{eq:P1}
P_{i\rightarrow\oplus}=\rho_{-}^{i}\equiv n^{i}_{-}(t)/[n^{i}_{+}(t)
+n^{i}_{-}(t)],
\end{equation}
and chooses $-$ with the probability
$P_{i\rightarrow\ominus}=1-P_{i\rightarrow\oplus}$, where $n^{i}_{+}(t)$
and $n^{i}_{-}(t)$, respectively, are the numbers of $+$ and $-$ neighbors
of $i$ at time step $t$, with $k_{i}=n^{i}_{+}(t)+n^{i}_{-}(t)$. The
expressions of probabilities, however, are valid only under the assumption
that the two resources have the \emph{same} accommodating capacity, i.e.,
$C_{+}=C_{-}$. In real-world resource allocation systems, typically we
have $C_{+} \neq C_{-}$. Consider, for example, the extreme case of
$C_{+} \gg C_{-}$. Suppose we have $\rho_{+}^{i}=\rho_{-}^{i}$ for agent
$i$. In this case, rationality demands a stronger preference to the
resource $+$ (i.e., with a higher probability). To investigate the
issues associated with the control of realistic Boolean dynamics, we define
\begin{eqnarray} \label{eq:interaction}
\begin{bmatrix} P_{i\rightarrow\oplus}\\ P_{i\rightarrow\ominus} \end{bmatrix}  = {\mathbb F} \cdot \hat{\rho_{i}} \equiv \begin{bmatrix} F_{+|+} & F_{+|-} \\F_{-|+} & F_{-|-} \end{bmatrix} \cdot \begin{bmatrix} \rho_{+}^{i} \\ \rho_{-}^{i} \end{bmatrix},
\end{eqnarray}
where $\mathbb{F}$ is the response function of each agent to its local
environment $\hat{\rho_{i}}$, i.e., the local neighbor's configuration with
$\rho_{+}^{i}=n_{+}^{i}(t)/k_i$ and $\rho_{-}^{i}=n_{-}^{i}(t)/k_i$. The
quantity $F_{r|r'}$ (or $F_{r|r}$) characterizes the contribution of the
$r'$-neighbors (or $r$-neighbors) to the probability for $i$ to adopt $r$.
The quantity $F_{r|r}$ represents the strength of \emph{assimilation} effect
among the neighbors, while $F_{r|r'}$ quantifies the \emph{dissimilation}
effect. Intuitively, the resource with a larger accommodating capacity would
have a stronger assimilation effect among agents. By definition, the
elements in each column in the matrix $\mathbb{F}$ satisfy
$F_{r|r'}+F_{r'|r'}=1$, i.e., the total probability for an agent to choose
$+$ and $-$ is unity.

Using the mean-field assumption that the configuration of neighbors is
uniform over the whole system, i.e., $\rho_{+}^{i}=\rho_{+}$, we have that
the stable solution for Eq.~(\ref{eq:interaction}) satisfies
$\hat{\bf{\rho}}={\mathbb F} \cdot \hat{\bf{\rho}}$, which leads to the
eigenstate of $\mathbb{F}$ as
\begin{eqnarray} \label{eq:optimal}
\hat{\rho}^{*}=\left(\begin{array}{c}
\rho_{+}^*\\\\ \rho_{-}^*
\end{array}\right)
=\left(\begin{array}{c}
\frac{F_{+|-}}{F_{+|-}+F_{-|+}}\\\\ \frac{F_{-|+}} {F_{+|-}+F_{-|+}}
\end{array}\right).
\end{eqnarray}
The rational response ($\mathbb{F}$) of agents to nonidentical
accommodation capacities of resources will lead to the equality
$\rho_{+}^*/\rho_{-}^*=C_+/C_-$, i.e., the stable fraction of the agent
densities in $+$ and $-$ is simply the ratio of the capacities. The
elements of $\mathbb{F}$ can then be defined accordingly using this
ratio and the condition $0\le F_{r|r'} \le 1$, which characterizes a
stronger preference to the resource with a larger capacity. For the specific
case of identical-capacity resources, we have $F_{+|-}=F_{-|+}$, and the
solution reduces to the result $\hat{\rho}^{*}=(0.5,0.5)^{\mathrm{T}}$ of the
original Boolean dynamics~\cite{ZHDHL:2013,ZWZYL:2005}. The optimal solution
for the resource allocation is $A^{*}=N \rho^{*}_{+} $.

A general measure of Boolean system's performance is the variance of $A_t$
with respect to the \emph{capacity} $C_{+}$:
\begin{equation} \label{eq:sigma}
\sigma^{2}=\frac{1}{T}\sum^{T}_{t=1}(A(t)-C_{+})^{2},
\end{equation}
which characterizes, over a time interval $T$, the statistical deviations
from the optimal resource utilization~\cite{ZWZYL:2005}. A smaller value
of $\sigma^{2}$ indicates that the resource allocation is more optimal.
A general phenomenon associated with Boolean dynamics is that, as agents
strive to join the minority group, an undesired herding behavior can emerge,
as characterized by large oscillations in $A(t)$. Our goal is to understand,
for the general setting of nonidentical resource capacities, the effect
of pinning control on suppressing/eliminating the herding behavior.

\paragraph*{\bf Pinning control scheme.}
Our basic idea to control the herding behavior is to ``pin'' certain
agents to freeze their states so as to realize optimal resource allocation,
following the general principle of pinning control of complex dynamical
networks~\cite{WC:2002,LWC:2004,CLL:2007,XLCCY:2007,TWF:2009,PF:2009,
YCL:2009,ZHDHL:2013}. Let $\rho_{p}$ be the fraction of agents to be
pinned, so the fraction of unpinned (or free) nodes is
$\rho_{f}=1-\rho_{p}$. The numbers of the two different types of agents,
respectively, are $N_{f}=N\cdot\rho_{f}$ and $N_{p}=N\cdot \rho_{p}$. The
free agents make choices according to local time-dependent information,
for whom the inputs from the pinned agents are fixed.

The two basic quantities characterizing a pinning control scheme are
the order of pinning (the way how certain agents are chosen to be
pinned) and the pinning pattern~\cite{ZHDHL:2013}. We adopt the
degree-preferential pinning (DPP) strategy in which the agents are
selected to be pinned according to their connectivity or degrees in
the underlying network. In particular, agents of higher degrees are
more likely to be pinned. This pinning strategy originated from the classic
control method to mitigate the effects of intentional attacks in complex
networks~\cite{AJB:2000,CNSW:2000,CEbAH:2001}. The selection of the
pinning pattern can be characterized by the fractions $\eta_{+}$ and
$\eta_{-}$ of the pinned agents that select $r=+$ and $r=-$, respectively,
where $\eta_{+}+\eta_{-}=1$. The quantities $\eta_{+}$ and $\eta_{-}$ are
thus the {\em pinning pattern indicators}. Different from the previous
work~\cite{ZHDHL:2013} that investigated the specific case of
$\eta_{+}=\eta_{-}=0.5$ (half-half pinning pattern), here we consider
the more general case where $\eta_{+}$ is treated as a variable. The
pinning schemes are implemented on random networks and scale-free
networks with different values of the scaling exponent $\gamma$ in the
power-law degree distribution~\cite{BA:1999,RRM:2005}
$P(k)\sim k^{-\gamma}$. As we will see below, one uniform optimal pinning
fraction $\rho_{p}$ exists for various values of the pinning pattern
indicator $\eta_{+}$.

\begin{figure}
\centering
\includegraphics[width=\linewidth]{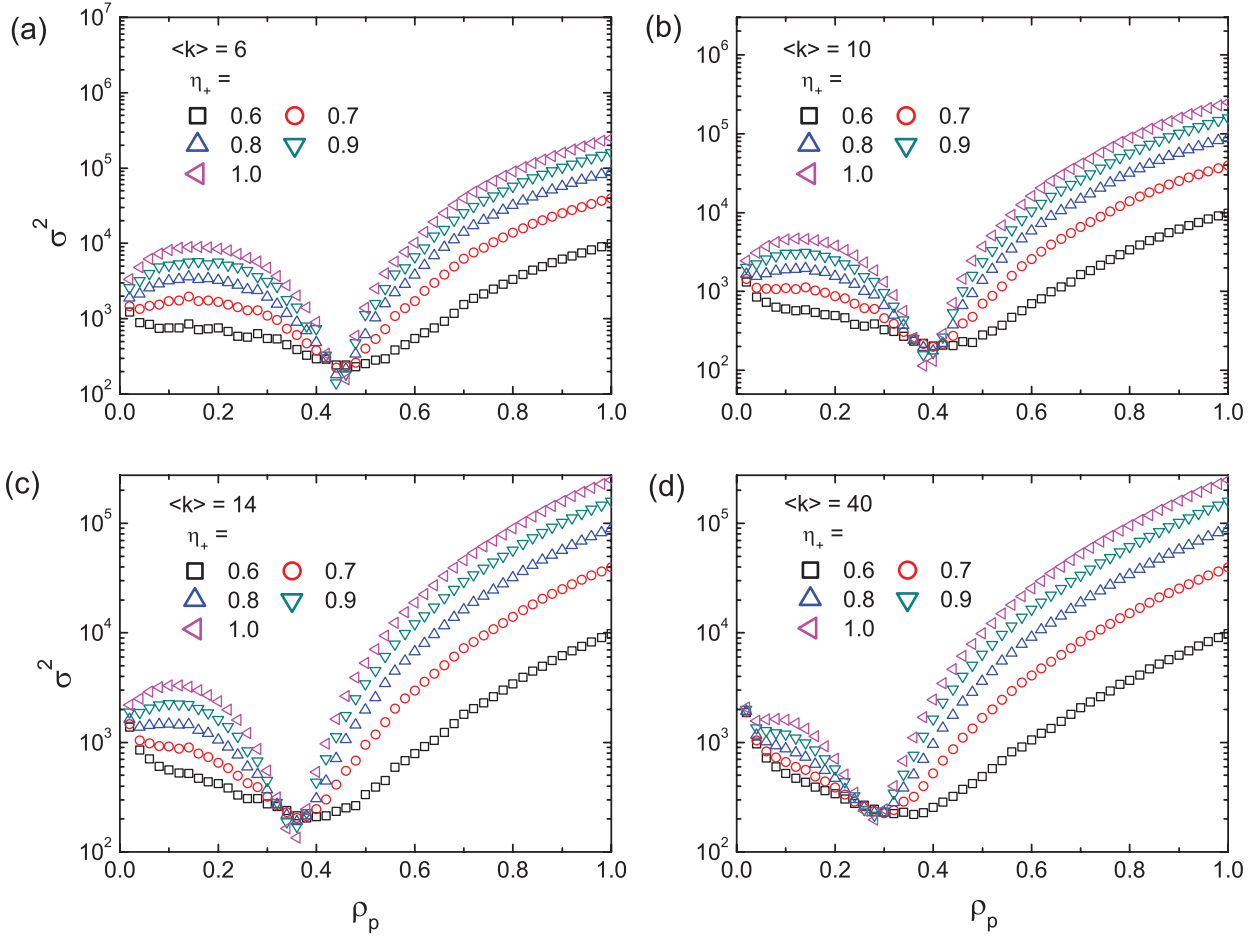}
\caption{\textbf{Variance $\sigma^2$ as a function of the pinning
fraction $\rho_{p}$ for scale-free networks of different connection
densities.} The average degree of the networks for simulation are
$\langle k\rangle=6$, $10$, $14$, to $40$ in (a-d), respectively, and the
value of the pinning pattern indicator $\eta_{+}$ ranges from $0.6$ to
$1.0$ for each panel. The results are averaged over $200$ realizations
for scale-free networks of size $N=1001$ and degree exponent $\gamma=3.0$.}
\label{fig:1}
\end{figure}

\begin{figure}
\centering
\includegraphics[width=0.9\linewidth]{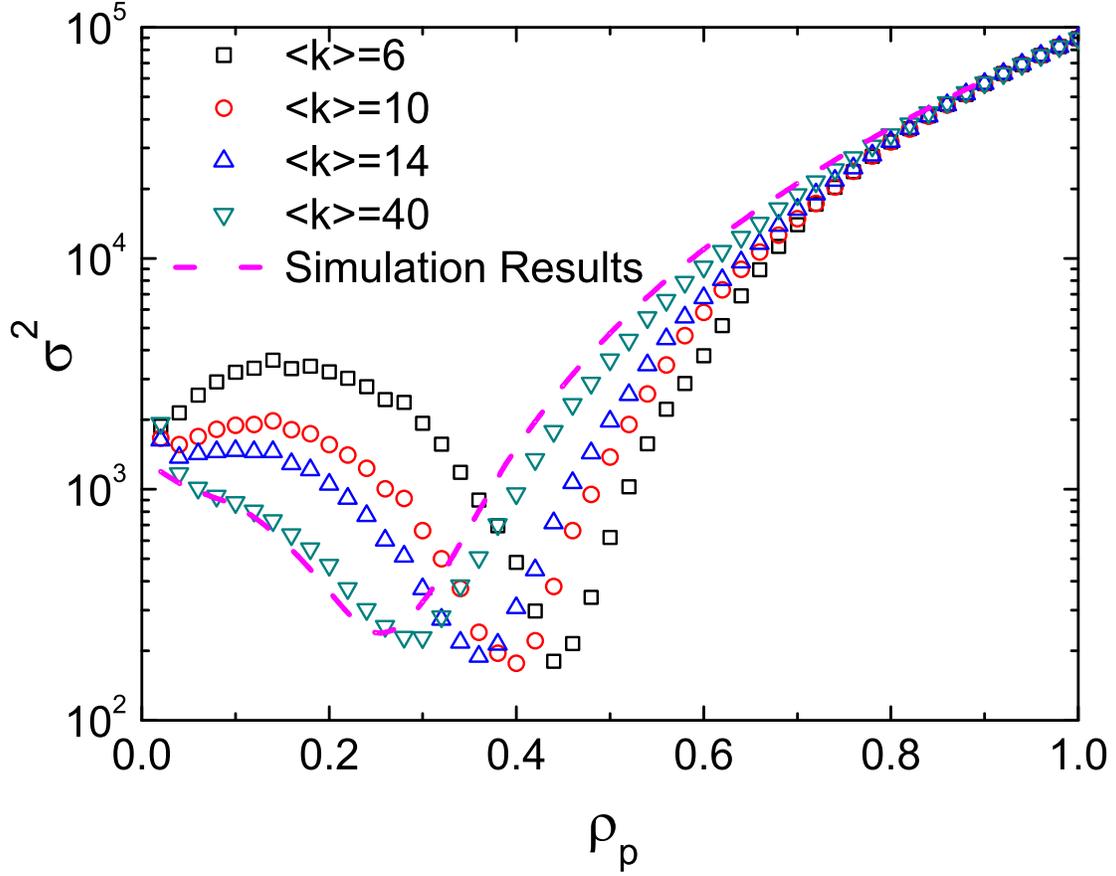}
\caption{\textbf{Theoretical prediction of the variance $\sigma^2$
in comparison with the simulation results.} The system has size $N=1001$
and power-law degree distribution $P(k)$ with scaling exponent $\gamma=3$.
The theoretical prediction does not depend on the value of the average
degree. In direct simulations, the values of the average degree are
$\langle k\rangle=6$, $10$, $14$, and $40$. In each case, the simulation
result is averaged over $200$ network realizations. The value of the
pinning pattern indicator is set to $\eta_{+}=0.8$}
\label{fig:4}
\end{figure}

\paragraph*{\bf Simulation Results.}
To gain insight, we first study the original Boolean dynamics with
$C_{+}/C_{-}=1$ and $\mathbb{F}=(0,1;1,0)$ for different values of the pinning
pattern indicator $\eta_{+}$. The game dynamics are implemented on scale-free
networks of size $N=1001$ and of the scaling exponent $\gamma=3.0$ with
the average degree $\langle k\rangle$ ranging from $6$ to $40$. The DPP
scheme is performed with pinning fraction $\rho_{p}$ and $\eta_{+}$
values ranging from $0.5$ (i.e., half-half pinning)
to $1.0$ (i.e., all to $+$ pinning). The variance $\sigma^2$ versus
$\rho_{p}$ for different values of $\eta_{+}$
and different degree $\langle k\rangle$ are shown in Fig.~{\ref{fig:1}}.
We see that, in general, systems with larger values of $\eta_{+}$ exhibit
larger variance, implying that a larger deviation of
$\eta_{+}$ from the ratio of the capacity $C_{+}/N$ can lead to
lower efficiency in resource allocation. Surprisingly, there exists
a universal optimal pinning fraction (denoted by $\rho^{*}_{p}$) about $0.4$,
where the variance $\sigma^2$ is minimized and exhibits an opposite trend
for $\rho_{p} > \rho^{*}_{p}$, i.e., larger values of $\eta_{+}$ result
in smaller values of $\sigma^2$. The implication is that, deviations
of $\eta_{+}$ from $C_{+}/N$ provide an opportunity to achieve better
performance (with smaller variances $\sigma^2$), due to the non-monotonic
behavior of $\sigma^2$ with $\rho_{p}$. To understand the
emergence of the optimal pinning fraction $\rho^{*}_{p}$,
we see from Fig.~{\ref{fig:1}} that the values of $\rho^{*}_{p}$
are approximately identical for different values of $\eta_{+}$, which
decrease with the average degree $\langle k\rangle$. As we will see below,
in the large degree limit $k\rightarrow \infty$, the value of $\sigma^2$
can be predicted theoretically (c.f., Fig.~\ref{fig:4}).

\begin{figure}
\centering
\includegraphics[width=\linewidth]{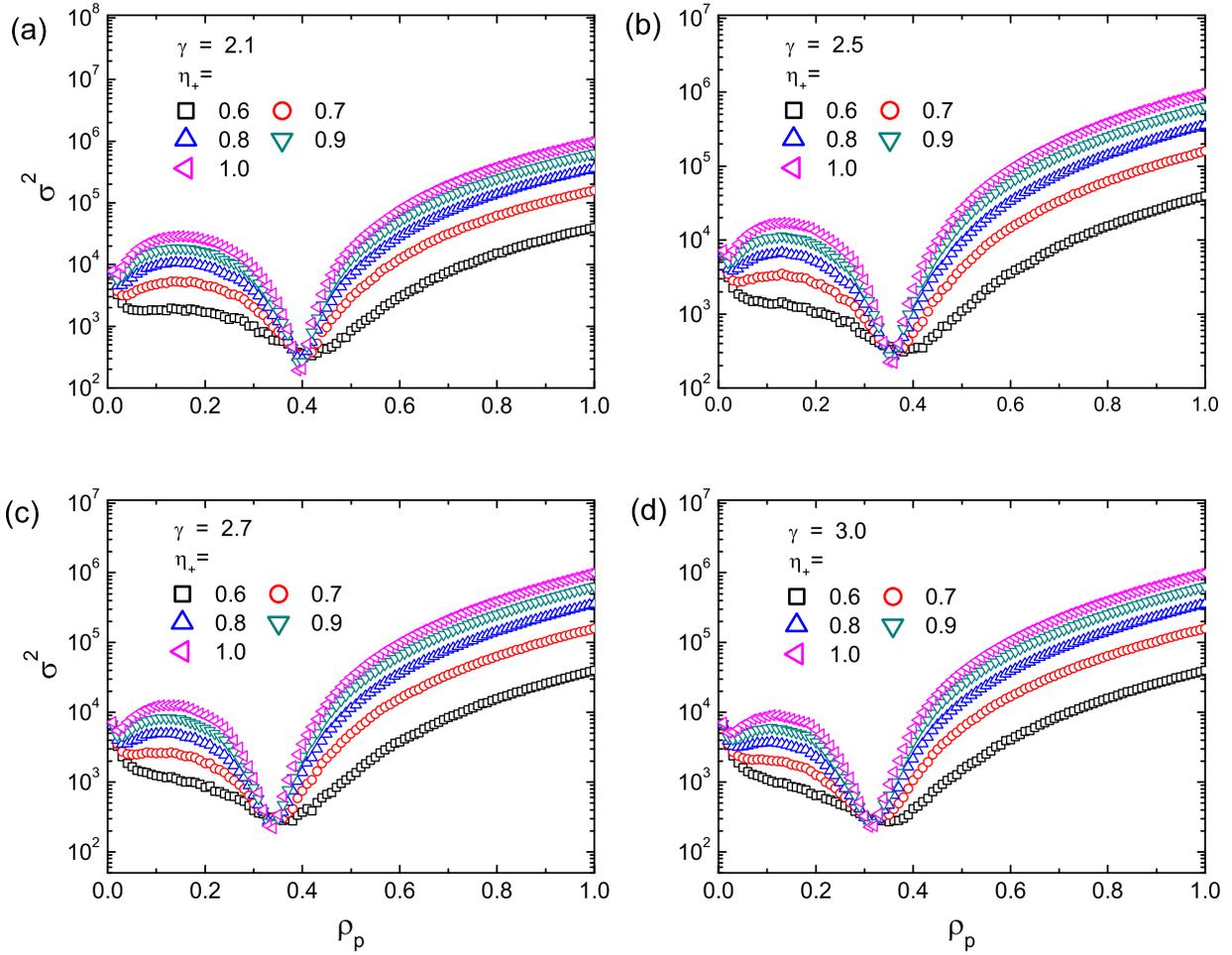}
\caption{\textbf{Variance $\sigma^2$ as a function of the pinning
fraction $\rho_{p}$ for scale-free networks of varying degrees of
heterogeneity.} The scaling exponents of the networks are
$\gamma=2.1$, $2.5$, $2.7$, and $3.0$ in (a-d), respectively, and
the value of the pinning pattern indicator $\eta_{+}$ ranges from $0.6$
to $1.0$ for each panel. The results are averaged over $200$ realizations
for scale-free networks of size $N=2001$ and average degree
$\langle k\rangle \approx 4$.}
\label{fig:2}
\end{figure}

\begin{figure}
\centering
\includegraphics[width=\linewidth]{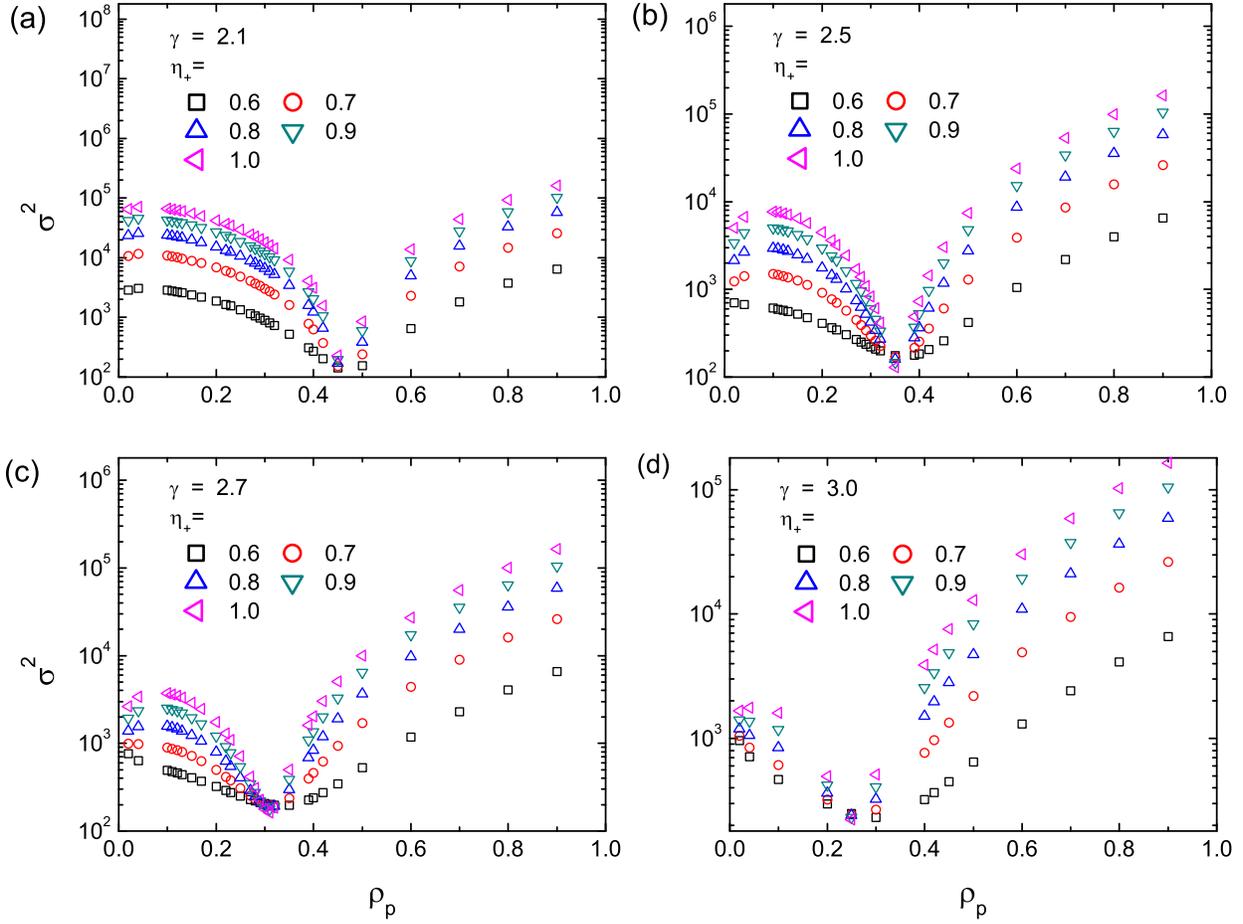}
\caption{\textbf{Theoretical prediction of variance $\sigma^2$ for
systems with different degree scaling exponents.} The system has size
$N=1001$ and power-law degree distribution $P(k)$ with different values
of the degree exponent: (a-d) $\gamma=2.1$, $2.5$, $2.7$, $3.0$,
respectively. In each case, the value of the pinning pattern indicator
$\eta_{+}$ ranges from $0.6$ to $1.0$.}
\label{fig:5}
\end{figure}

Simulations using scale-free networks of different degrees of
heterogeneity also indicate the existence of the universal optimal
pinning control strategy, as can be seen from the behaviors of the
variance calculated from scale-free networks of different degree
exponents (Fig.~{\ref{fig:2}}), where smaller values of $\gamma$
point to a stronger degree of heterogeneity of the system. We see
that an optimal value of $\rho_{p}^*$ exists for all cases,
which decreases only slightly with $\gamma$, i.e., more heterogeneous
networks exhibit larger values of the optimal pinning fraction
$\rho_{p}^*$, a phenomenon that can also be predicated theoretically
(c.f., Fig.~\ref{fig:5}). \\
\\ \noindent
{\large\textbf{Theoretical Analysis}} \\
\\ \noindent
The phenomenon of the existence of a universal optimal pinning fraction
$\rho^*_{p}$, independent of the specific values of pinning pattern
indicator $\eta_{+}$, is remarkable. Here we develop a quantitative
theory to explain this phenomenon.

\begin{figure}
\centering
\includegraphics[width=\linewidth]{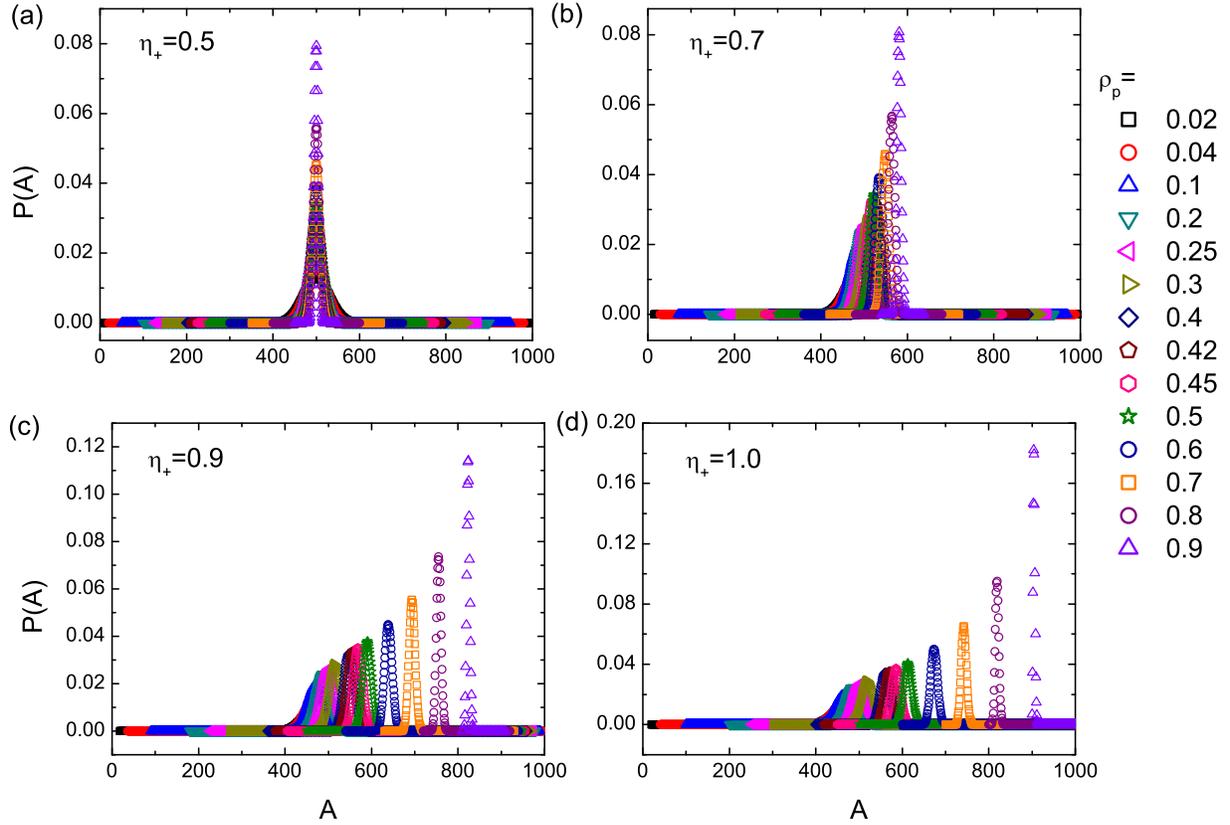}
\caption{\textbf{Theoretical prediction of the probability density
distribution of attendance \emph{A}.} The distribution $P(A)$ is
obtained from the transition matrix Eq.~(\ref{eq:T20pin2}) for
$N=1001$. The value of the pinning pattern indicator $\eta_{+}$ is
set as $0.5$, $0.7$, $0.9$ and $1.0$ in (a-d), respectively, and the
pinning fraction $\rho_{p}$ ranges from $0.02$ to $0.9$.}
\label{fig:3}
\end{figure}

To begin, we note that MG is effectively a stochastic dynamical process
due to the randomness in the selection of strategies by the agents.
The variance of the system, a measure of the efficiency of the system,
is determined by two separated factors. The first, denoted as
$X_1(\delta)$, is the intrinsic fluctuations of $A$ about its expected
value $A^{*}$, defined as $\delta\equiv \sqrt{\langle(A-A^*)^2\rangle}$,
which can be calculated once the stable distribution of attendance $P(A)$
is known, where $P(A)$ can be obtained either analytically (c.f.,
Fig.~\ref{fig:3}) or numerically. The second factor, denoted as
$X_{2}(\varepsilon)$, is the difference of the expected value $A^{*}$
from the capacity $C_{+}$ of the system: $\varepsilon \equiv A^{*}-C_{+}$,
which leads to a constant contribution to the variance of the system.
Taking into account the two factors, we can write the system
variance $\sigma^2$ [defined in Eq.~(\ref{eq:sigma})] as
\begin{eqnarray} \label{eq:sigmaF1F2}
\sigma^2\equiv\langle(A-C_{+})^2\rangle =
\langle(A-A^*+A^*-C_{+})^2\rangle =\delta^2+\varepsilon^2,
\end{eqnarray}
a direct summation of the two factors $X_{1}(\delta)=\delta^2$ and
$X_{2}(\varepsilon)=\varepsilon^2$. In contrast to the simplified case
discussed in the previous works~\cite{ZWZYL:2005,ZHDHL:2013}, the expected
value $A^{*}$ in the dynamical process is not necessarily equal to the
capacity $C_{+}$. Nonzero values of $\varepsilon$ are a result of
biased pinning pattern ($\eta_{+}\neq 0.5$) or improper response
to the capacities of resources. In addition, recent studies of flux-fluctuation law in complex systems also found that, the variance of the system is determined by the two factors: the intrinsic fluctuation and the systematic external drives~\cite{deMB:2004,DA:2006,YYK:2007,MGLM:2008,ZHYXW:2010,ZHHLYX:2013,HDHL:2014}.

\paragraph*{\bf Stable distribution of attendance.}
To quantify the process of biased pinning control, we derive a
discrete-time master equation and then discuss the effect of network
topology on control.

\paragraph*{Discrete-time master equation for biased pinning control.}
To understand the response of the Boolean dynamics to pinning control with
varied values of the pinning pattern indicator $\eta_{+}$, we generalize
our previously developed analysis~\cite{ZHDHL:2013}. Let $P_{p}$ be the
probability for a \emph{neighbor} of one given \emph{free} agent to be
pinned so that the probability of encountering a free agent is $P_f=1-P_p$.
The transition probability of the system from $A(t)$ to $A(t+1)$ can be
expressed in terms of $\eta_{+}$. In particular, note that the state transition
is due to updating of the $N_f$ free agents, as the remaining $N_{p}$
agents are fixed. To simplify notations, we set $A(t)=i$, $A(t+1)=k$,
and $A(t+2)=j$, for $i,k,j\in [0,N]$. The conditional transition
probability from $i$ at $t$ to $k$ at $t+1$ is
\begin{eqnarray} \label{eq:T_onestep}
P(k,t+1|i,t)=&&{N_f\choose k-N_{p}\eta_{+}}\times (P_{p}\eta_{-}+P_{f}
\frac{N_f-(i-N_p\eta_{+})}{N_f})^{k-N_{p}\eta_{+}} \nonumber\\
&&\times (P_{p}\eta_{+}+P_{f}\frac{i-N_{p}\eta_{+}}{N_{f}})^{N_{f}
-(k-N_{p}\eta_{+})},
\end{eqnarray}
where $P_{p}\eta_{-}+P_{f}[N_f-(i-N_p\eta_{+})]/N_f$ is the probability
for a free agent to choose $+$ with the first and second terms
representing the contributions of the pinned $-$ and free $-$ neighbors,
respectively. In the Boolean system, the values of attendance $A$ oscillate
about its equilibrium value~\cite{ZHDHL:2013}. The transition probability
between the state at $t$ and $t+2$ can be expressed as a function of
$\eta_{+}$:
\begin{eqnarray} \label{eq:T20pin2}
T(j,i)\equiv&&P(j,t+2|i,t)=\sum_{k} P(j,t+2|k,t+1) \cdot
P(k,t+1|i,t) \nonumber\\
=&&\sum_{k}\{ [{N_f\choose k-N_{p}\eta_{+}}\times (P_{p}\eta_{-}+P_{f}\frac{N_f-i+N_p\eta_{+}}{N_f})^{k-N_{p}\eta_{+}} \nonumber\\
&&\times (P_{p}\eta_{+}+P_{f}\frac{i-N_p\eta_{+}} {N_f})^{N_{f}-(k-N_{p}\eta_{+})} ] \nonumber\\
&&\times [ {N_f\choose j-N_{p}\eta_{+}} \times (P_{p}\eta_{-}+P_{f}\frac{N_f-k+N_p\eta_{+}}{N_f})^{j-N_{p}\eta_{+}} \nonumber\\
&&\times (P_{p}\eta_{+}+P_{f}\frac{k-N_p\eta_{+}} {N_f})^{N_{f}-(j-N_{p}\eta_{+})} ] \}.
\end{eqnarray}
Equation~(\ref{eq:T20pin2}) takes into account the effect of pinning
patterns, which was ignored previously~\cite{ZHDHL:2013}. The
resulting balance equation governing the dynamics of the Markov chains
becomes
\begin{eqnarray} \label{eq:distribution}
P(j)=\sum_{i}P(j,t+2|i,t)P(i) \equiv \sum_{i}T(j,i)P(i),
\end{eqnarray}
which is the discrete-time master equation. The stable state that the
system evolves into can be defined in the matrix form as
\begin{eqnarray} \label{eq:distribution2}
{\bm{P}}_{A}={\mathbb{T}}{\bm P}_{A},
\end{eqnarray}
where $\mathbb{T}$ is an $(N+1)\times (N+1)$ matrix with elements
$\mathbb{T}_{ji}=T(j,i)$, and ${\bm{P}}_{A}$ is the corresponding
vector of $P(A)$ with $A$ ranging from $0$ to $N$.

The probability distribution $P(A)$ is a binomial function with
various expectation values, as shown in Fig.~{\ref{fig:3}}. In
addition, the probability $P(A)$ is zero for
$A \in [0,N_p\eta_{+}] \cup [N_p(1-\eta_{+}),N]$, which defines
the boundary condition in the sense that there are $N_p=N\cdot\rho_{p}$
pinned agents. Once the stable distribution $P(A)$ is obtained
from Eq.~(\ref{eq:distribution2}), the cumulative variance of the
system can be calculated from
\begin{eqnarray} \label{eq:sigmaA}
\sigma^2=\sum_{A=0}^N P(A)(A-C_{+})^2.
\end{eqnarray}
The theoretical prediction of $\sigma^{2}$ as a function of $\rho_{p}$
can thus be made through ({\em a}) identifying the function $P_{p}(\rho_{p})$,
({\em b}) defining the matrix $\mathbb{T}$ that depends on $\eta_+$ and
$P_{p}(\rho_{p})$, and ({\em c}) calculating the stable state $P(A)$.

\paragraph*{Effect of network topology on pinning control.}
The topology of the network system has an effect on the probability $P_{p}$.
For the particular case of scale-free networks with degree exponent
$\gamma=3$, our previous work~\cite{ZHDHL:2013} demonstrated that
preferential pinning of the large-degree agents leads to
$P_{p}=\sqrt{\rho_{p}}$. Here, we consider systems with degree distribution
$P(k)=(\gamma-1)/(k_{\mathrm{min}}^{1-\gamma}k^{\gamma})$,
where $k_{\mathrm{min}}$ is the minimum degree of the network. For the
DPP scheme where pinning occurs in the order from large to small degree
agents, the relation between the minimum degree of \emph{pinned} agents
(denoted by $k^{\prime}$) and the pinning fraction $\rho_{p}$ is
\begin{eqnarray} \label{eq:rhop{k}}
\rho_{p}(k^{\prime})=\int_{k^\prime}^{k_{\mathrm{max}}}P(k)dk.
\end{eqnarray}
For a given pinning fraction $\rho_{p}(k^{\prime})$ in which all the
agents with $k>k'$ are pinned, the probability $P_{p}$ for one neighbor
of a given \emph{free} agent to be a pinned agent is given by
\begin{eqnarray} \label{eq:rho_k_{Pk}}
P_{p}(k^{\prime})=\frac{\int_{k^{\prime}}^{k_{\mathrm{max}}}kP(k)dk}
{\int_{k_{\mathrm{min}}}^{k_{\mathrm{max}}}kP(k)dk}.
\end{eqnarray}
Equations~(\ref{eq:rhop{k}}) and (\ref{eq:rho_k_{Pk}}) are applicable
to DPP scheme on networks of any degree distribution $P(k)$. For a
scale-free network, $P_{p}$ as a function of $\rho_{p}$ can be expressed as
\begin{eqnarray} \label{eq:pfp_rho}
P_{p}(\rho_{p})={(\frac{k^{\prime}}{k_{\mathrm{min}}})}^{2-\gamma}
={\rho_{p}}^{\frac{2-\gamma}{1-\gamma}}.
\end{eqnarray}
For the special case of $\gamma=3$, Eq.~(\ref{eq:pfp_rho}) reduces to
the specific relationship obtained earlier~\cite{ZHDHL:2013}:
$P_{p}(\rho_{p})=\sqrt{\rho_{p}}$. As indicated by
Eqs.~(\ref{eq:T20pin2}-\ref{eq:sigmaA}), the specific form of matrix
$\mathbb{T}$ with respect to $\rho_{p}$ can be obtained by substituting
Eq.~(\ref{eq:pfp_rho}) into Eq.~(\ref{eq:T20pin2}), leading to the
distribution $P(A)$ and finally the variance of the system $\sigma^{2}$
as a function of $\rho_{p}$. Figure~\ref{fig:4} displays the theoretical
predicted $\sigma^{2}$ (dashed curves) for various values of the pinning
fraction $\rho_{p}$ and of the pinning pattern indicator $\eta_{+}$. The
trend and, more importantly, the existence of the optimal pinning fraction
$\rho_{p}^*$, agree well with the simulation results (marked with
different symbols). In the limit $\langle k \rangle\rightarrow\infty$,
the system approaches a well-mixed state that can be fully characterized
by Eq.~(\ref{eq:pfp_rho}), indicating that the simulation results
approach the curve predicted by the mean-field theory as the average
degree $\langle k\rangle$ is increased.

Figure~\ref{fig:5} shows the theoretical prediction of $\sigma^{2}$ for
scale-free networks with different values of the degree exponent $\gamma$,
which agrees well with the results from direct simulation as in
Fig.~\ref{fig:2}. For the case of highly heterogeneous networks ($\gamma=2.1$),
the theoretical prediction deviates slightly from the numerical results
for the reason that the networks in simulation inevitably exhibit certain
topological features that are not taken into account in the theoretical
analysis of $P_{p}(\rho_{p})$, such as the degree associativity.

\paragraph*{\bf Optimal pinning.}
Our analysis based on the master equation (\ref{eq:distribution}) applies
to systems with $F_{+|-}=F_{-|+}=1$ and identical resource capacity. We
now consider the more general case of varying $\mathbb{F}$ values to
further understand the optimal pinning control scheme.

\paragraph*{Deviation of expected attendance from resource capacity.}
The dependence of $A^{*}$ on $\eta_{+}$ can be obtained through the
general form of the response matrix $\mathbb{F}$. For convenience, we use
the column vector ${\hat{\eta}}=(\eta_{+},\eta_{-})^{\mathrm{T}}$ to
denote the fraction of the agents pinned at $+$ and $-$, where
$\eta_{+}+\eta_{-}=1$, ${\hat{\omega}}=(\omega_{+},\omega_{-})^{\mathrm{T}}$
is the fraction of free agents adopting strategies $+$ and $-$,
respectively, with $\omega_{+}+\omega_{-}=1$. The state of the system
can be expressed as $\hat{\rho}(t)= \rho_{p}{\hat{\eta}} +
\rho_{f}{\hat{\omega}}(t)$, from which we have
\begin{eqnarray} \label{eq:EtaRhoP}
{\hat{\omega}}(t)=[\hat{\rho}(t)-\rho_{p}{\hat{\eta}}]/\rho_{f},
\end{eqnarray}
At the next time step, the expected value of the state based on
$\hat{\rho}(t)$ through the response matrix $\mathbb{F}$ can be written as
\begin{eqnarray} \label{eq:rhot+1}
\hat{\rho}(t+1)=\rho_{p}{\hat{\eta}}+\rho_{f}[P_{p}{\mathbb{F}}{\hat{\eta}}
+P_{f}{\mathbb{F}}{\hat{\omega}}(t)].
\end{eqnarray}
Substituting Eq.~(\ref{eq:EtaRhoP}) into Eq.~(\ref{eq:rhot+1}), we get
the relationship between $\hat{\rho}(t+1)$ and $\hat{\rho}(t)$. A
self-consistency process stipulated by Eqs.~(\ref{eq:EtaRhoP}) and
(\ref{eq:rhot+1}) can yield the stable state of the system with the
expected number of agents choosing $+$ given by
\begin{eqnarray} \label{eq:pstabel}
A^{*}=N\frac{F_{+|-}(\rho_{p}-1)-\rho_{p} \eta_{+}(F_{+|-}+F_{-|+})
+\eta_{+}P_{p}(F_{+|-}+F_{-|+}-1)}{P_{p}(F_{+|-}+F_{-|+}-1)-(F_{+|-}+F_{-|+})}.
\end{eqnarray}
In a free system without pinning, the rational response $\mathbb{F}$ of
agents to nonidential capacities of resources leads to Eq.~(\ref{eq:optimal}),
implying the relationship $C_{+}=NF_{+|-}/(F_{+|-}+F_{-|+})$. From
Eq.~(\ref{eq:pstabel}), we can obtain $\varepsilon$ as a function of the
value of the pinning pattern indicator $\eta_{+}$, the elements of the
matrix $\mathbb{F}$, the pinning fraction $\rho_{p}$, and the parameter
$P_{p}$ associated with network topology. We have
\begin{eqnarray} \label{eq:deltap}
\varepsilon= A^{*}-C_{+}= && N\cdot[F_{+|-}(\eta_{+}-1)
+F_{-|+}\eta_{+}]\cdot \nonumber\\
&&\frac{[\rho_{p}(F_{+|-} +F_{-|+})-P_{p}(F_{+|-}+F_{-|+}-1)]}{(F_{+|-}
+F_{-|+})\cdot[F_{+|-}+F_{-|+}-(F_{-|+}+F_{-|+}-1)P_{p}]},
\end{eqnarray}
which has the form of separated variables associated with $\eta_{+}$
and $\rho_{p}$.

\paragraph*{Optimal pinning pattern and fraction.}
Optimizing the system requires minimum $X_{2}(\varepsilon)$, i.e.,
$\varepsilon=0$ in Eq.~(\ref{eq:deltap}), leading to two independent
solutions:
\begin{subequations}
\begin{eqnarray}
\frac{\eta_{+}^{*}}{\eta_{-}^{*}}&=&\frac{F_{+|-}}{F_{-|+}}
=\frac{C_{+}}{C_{-}}, \label{opt_pin_pattern} \\
P_{p}(\rho_{p}^*)&=&\frac{F_{+|-}+F_{-|+}}{F_{+|-}+F_{-|+}-1}\rho_{p}^{*},
\label{eq:optimal_rho}
\end{eqnarray}
\end{subequations}
which respectively correspond to the optimal value of the pinning pattern
indicator $\eta_{+}^{*}$ and the optimal pinning fraction $\rho_{p}^{*}$.
Here, for convenience, we define a parameter:
$\beta\equiv (F_{+|-}+F_{-|+})/(F_{+|-}+F_{-|+}-1)$ so that
Eq.~(\ref{eq:optimal_rho}) can be expressed concisely as
$P_{p}(\rho_{p}^*)=\beta\cdot\rho_{p}^{*}$. Once the values of $\eta_{+}$
and $\rho_{p}$ satisfy either Eq.~(\ref{opt_pin_pattern}) or
Eq.~(\ref{eq:optimal_rho}), we can obtain $X_{2}(\varepsilon)=0$. The
variance $\sigma^{2}$ depends on the fluctuation factor $X_{1}(\delta)$ only.

Equation~({\ref{opt_pin_pattern}}) specifies the pinning pattern with
the same ratio as that of the resource capacity. The Boolean dynamics studied
previously~\cite{ZHDHL:2013} is a special case where the optimal pinning
pattern indicator is ${\eta_{+}^*}/{\eta_{-}^*}=1$ for systems with
$C_+=C_-$, and the variance $\sigma^2$ is simply determined by the factor
$X_{1}(\delta)$.

\begin{figure}
\centering
\includegraphics[width=\linewidth]{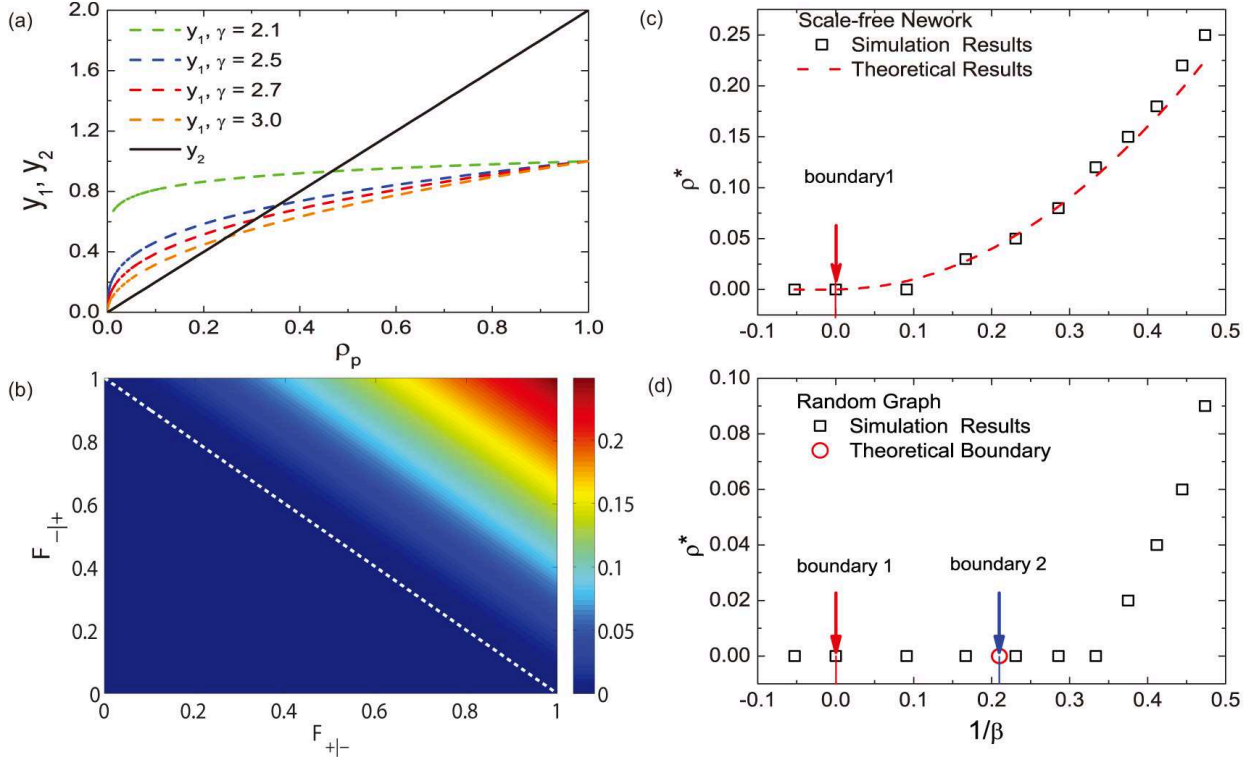}
\caption{\textbf{Optimal pinning fraction.} (a) Intersections of the
curves $y_1=P_{p}(\rho_{p})$ and $y_2=\beta\cdot\rho_{p}$ denote
nonzero optimal pinning fraction $\rho_{p}^{*}$ given by
Eq.~(\ref{eq:optimal_rho}). The scale-free networks have the degree
exponents $\gamma=2.1$, $2.5$, $2.7$, and $3.0$, respectively. The
response function is for $F_{+|-}=F_{-|+}=1$ (corresponding to
$C_{+}=C_{-}$). (b) Contour map of $\rho_{p}^*$ in the parameter space
of $F_{+|-}$ and $F_{-|+}$ for scale-free networks with $\gamma=3$. In
the lower-left region below the boundary $F_{+|-}+F_{-|+}=1$ (white
dashed line), nonzero solution of $\rho_{p}^*$ cannot be obtained.
(c) Optimal pinning fraction $\rho_{p}^{*}$ as a function of $1/\beta$ for
scale-free networks. The analytical results from Eq.~(\ref{eq:optimal_rho})
(red solid curve) and the simulation results (black open squares) agree
well with each other. The red arrow marks the theoretical prediction of
the boundary, where nonzero $\rho_{p}^*$ solutions exist on the left side.
(d) For ER random networks, $\rho_{p}^{*}$ as a function of $1/\beta$.
Theoretical results from Eq.~(\ref{eq:optimal_rho}) (red open circle)
and simulation results (black open squares) are shown. The boundaries
$1$ and $2$ obtained theoretically (pointed to by solid arrows),
respectively, stand for the constraint in Eqs.~(\ref{eq:re_func_condition})
and (\ref{eq:disc_net_condition2}). In (c) and (d), the value of $F_{-|+}$
varies but $F_{+|-}$ is set to $0.9$. The scale-free and random networks
used in the simulations have $\langle k \rangle=40$ and $N=2001$.}
\label{fig:6}.
\end{figure}

From Eq.~(\ref{eq:optimal_rho}), we see that the optimal pinning fraction
$\rho_{p}^*$ is independent of $\eta_{+}$ but depends on both the network
structure through $P_{p}(\rho_{p})$ and on the response function $\mathbb{F}$.
Additionally, the condition $\rho_{p}\in [0,1]$ and nonzero denominator
require
\begin{equation} \label{eq:re_func_condition}
F_{+|-}+F_{-|+}-1>0.
\end{equation}
The function $P_{p}(\rho_{p})$ for scale-free networks, as in
Eq.~(\ref{eq:pfp_rho}), increases monotonically with $\rho_{p}$.
Figure~\ref{fig:6}(a) displays the curves $y_1=P_{p}(\rho_{p})$
and $y_2=\beta\cdot\rho_{p}$, i.e., both sides of
Eq.~(\ref{eq:optimal_rho}). The existence of nonzero $\rho_{p}^*$
for $y_1=y_2$ demands
\begin{equation} \label{eq:cont_net_conditon}
\frac{dP_{p}(\rho_{p})}{d\rho_{p}}|_{\rho_{p}=0} > \beta.
\end{equation}
For scale-free networks, $P_{p}(\rho_{p})$ diverges at $\rho_{p}=0$.
Equation~(\ref{eq:cont_net_conditon}) thus holds, implying that
the DPP pinning scheme has a nonzero optimal pinning fraction
$\rho_{p}^*$, leading to $\varepsilon=0$. However, for homogeneous
networks, Eq.~(\ref{eq:cont_net_conditon}) may not hold. In this
case, a more specific implicit condition can be obtained from
Eq.~(\ref{eq:cont_net_conditon}) through the following analysis.
In particular, without an analytical expression of $P_{p}(\rho_{p})$,
the derivative of $P_{p}(\rho_{p})$ with respect to $\rho_{p}$ can
be obtained from Eqs.~(\ref{eq:rhop{k}}) and (\ref{eq:rho_k_{Pk}}):
\begin{eqnarray} \label{eq:disc_net_condition}
\frac{dP_{p}(\rho_{p})}{d\rho_{p}}=&&\frac{dP_{p}(k^{\prime})}
{dk^{\prime}}\cdot \frac{dk^{\prime}}
{d\rho_{p}(k^{\prime})}= \frac{k^{\prime}P(k^{\prime})}
{\langle k\rangle P(k^{\prime})}=\frac{k^{\prime}}
{\langle k \rangle}.
\end{eqnarray}
For degree preferential pinning, in the limit $\rho_{p}\rightarrow 0$,
the maximum degree for \emph{free} agents is
$k^{\prime}\rightarrow k_{\mathrm{max}}$. We thus have
\begin{eqnarray} \label{eq:disc_net_condition2}
\frac{dP_{p}(\rho_{p})}{d\rho_{p}}|_{\rho_{p}=0} =\frac{k_{\mathrm{max}}}
{\langle k \rangle}>\beta,
\end{eqnarray}
which requires that the network be heterogeneous. For $F_{+|-}=F_{-|+}=1$,
we have ${k_{\mathrm{max}}}/{\langle k \rangle}>2$, ensuring the
existence of a nonzero $\rho_{p}^*$ value for $\varepsilon=0$.

The contour map of the optimal pinning
fraction $\rho_{p}^*$ in the parameter
space of $F_{+|-}$ and $F_{-|+}$ for scale-free networks with $\gamma=3$
is shown in Fig.~\ref{fig:6}(b). The boundary $F_{+|-}+F_{-|+}=1$
associated with condition Eq.~(\ref{eq:re_func_condition}) is represented
by the white dashed line, where nonzero solutions of $\rho_{p}^*$ do not
exist below the lower-left region.
Figures~\ref{fig:6}(c) and~\ref{fig:6}(d) show $\rho_{p}^*$ for $\varepsilon=0$ as a function of $1/\beta$
for scale-free and random networks, respectively, where $F_{-|+}$ is
varied and $F_{+|-}$ is fixed to $0.9$. The theoretical prediction of
$\rho_{p}^*$ [red solid curve in (c) and red open circle in (d)]
is given by the intersections of the curves $y_1$ and $y_2$ in
Fig.~\ref{fig:6}(a). For scale-free networks, since
Eq.~(\ref{eq:cont_net_conditon}) holds,
Eq.~(\ref{eq:re_func_condition}) is the only constraint on the value
of $1/\beta$ (red dashed arrow), with the region at the right-hand side
yielding nonzero $\rho_{p}^*$ solutions. The red solid curve in
Fig.~\ref{fig:6}(c) represents the theoretical prediction, and the
open squares denote the simulation results from scale-free networks
of size $N = 2001$, power-law exponent $\gamma=3$, and average degree
$\langle k\rangle=40$.

For random networks, the existence of nonzero $\rho_{p}^*$ solutions
requires that Eqs.~(\ref{eq:re_func_condition}) and
(\ref{eq:disc_net_condition}) or~(\ref{eq:disc_net_condition2}) hold.
For the Poisson degree distribution, the maximum degree of the network
can be calculated from
\begin{eqnarray} \label{eq:er_max_k}
\frac{e^{-k_{\mathrm{max}}}\langle k \rangle}{k_{\mathrm{max}}!}
\approx \frac{1}{N}.
\end{eqnarray}
We can obtain an estimate of the value of $1/\beta$ that satisfies
Eq.~(\ref{eq:disc_net_condition2}), as indicated by the blue arrow
(labeled as boundary 2) in Fig.~\ref{fig:6}(d). The right-hand side
of this point satisfies both Eqs.~(\ref{eq:re_func_condition}) and
(\ref{eq:disc_net_condition2}), implying the existence of nonzero
$\rho_{p}^*$. Comparison of the results from random and scale-free
networks with different scaling exponents (Figs.~\ref{fig:2},
~\ref{fig:5} and~\ref{fig:6}) shows that, stronger heterogeneity
tends to enhance the values of $\rho_{p}^*$, which can also be seen from
Eq.~(\ref{eq:cont_net_conditon}).

To better understand the non-monotonic behavior of $\sigma^2$ with
$\rho_{p}$, we provide a physical picture of the behavioral
change for $\rho_{p}$ greater or less than $\rho_{p}^*$. The effect of
pinning control is determined by the number of edges between pinned and
free agents, which are \emph{pinning-free edges}. For a small pinning
fraction $\rho_{p}$, the average effect per pinned agent on the system
(represented by the number of pinned-free edges per pinned agent) is
relatively large. However, as $\rho_{p}$ is increased, the average impact is
reduced for two reasons: ({\em a}) an increase in the edges within the pinned
agents' community itself (i.e., two connected pinned agents), which
has no effect on control, and ({\em b}) a decrease in the number of free
agents, which directly reduces the number of pinned-free edges. Consider the
special case of $\eta_{+}=1$ and $C_{+}=C_{-}$. For small $\rho_{p}$, the
pinned $+$ agents have a significant impact so that the free agents
tend to overestimate the probability of winning by adopting $-$. In
this case, the expected value $A^*$ is smaller than $0.5 N$, corresponding
to $\varepsilon<0$. For highly heterogeneous systems, the average impact
per pinned agent is larger for a given small value of $\rho_p$. As
$\rho_{p}$ is increased, the average influence per pinned agent reduces
and, consequently, $A^*$ restores towards $0.5N$. For $A^*=0.5N$ and
$\varepsilon=0$, the system variance [Eq.~(\ref{eq:sigmaF1F2})] is
minimized due to $X_{2}(\varepsilon)=0$, and the corresponding pinning
fraction achieves the optimal value $\rho_{p}^*$. For strongly
heterogeneous systems, due to the large initial average impact caused
by pinning the hub agents, the optimal pinning fraction $\rho_{p}^*$
appears in the larger $\rho_p$ region. Further increase in $\rho_{p}$
with $\eta_{+}=1$ will lead to $\varepsilon>0$ and $A^*>0.5N$, thereby
introducing nonzero $X_{2}(\varepsilon)$ again and, consequently,
generating an increasing trend in $\sigma^2$.

\paragraph*{Collapse of variance.} For certain networks, the variance
$\sigma^{2}(\eta_{+},\rho_{p})$ is determined by the values of the pinning
pattern indicator $\eta_{+}$ and the pinning fraction $\rho_{p}$. Our
analysis so far focuses on the contribution of $X_{2}(\varepsilon)$ to
the variance $\sigma^2$ as the pinning fraction $\rho_{p}$ is increased
but for fixed $\eta_{+}$. It is thus useful to define a quantity related
to the variance $\sigma^2$, which can be expressed in the form of
separated variables. For two different values of the pinning pattern
indicator, $\eta_{+}$ and $\eta_{+}^{\prime}$, for a given value of
$\rho_{p}$, the relative weight of $X_{2}(\varepsilon)$ can be obtained from
Eq.~(\ref{eq:deltap}) as
\begin{eqnarray} \label{eq:lambda}
\lambda(\eta_{+},\eta_{+}^{\prime},\rho_{p})  \equiv
\frac{X_{2}(\varepsilon(\eta_{+},\rho_{p}))}{X_{2}
(\varepsilon(\eta_{+}^{\prime},\rho_{p}))}
=[\frac{\varepsilon(\eta_{+},\rho_{p})}
{\varepsilon(\eta_{+}^{\prime},\rho_{p})}]^2
=[\frac{F_{+|-}(\eta_{+}-1)+F_{-|+}\eta_{+}}
{F_{+|-}(\eta_{+}^{\prime}-1)+F_{-|+}\eta_{+}^{\prime}}]^2,
\end{eqnarray}
where $\varepsilon(\eta_{+},\rho_{p})$ is a function of both $\eta_{+}$
and $\rho_{p}$. Remarkably, the ratio $\lambda$ depends on $\eta_{+}$
and $\eta_{+}^{\prime}$ but it is independent of $\rho_{p}$, due to the
form of separated variables in Eq.~(\ref{eq:deltap}). From the simple
relationship Eq.~(\ref{eq:lambda}), we can define the relative changes
in these quantities due to an increase in the value of $\eta_{+}$
from a \emph{reference value $\eta_{+}^{\prime}$} as
\begin{eqnarray}
&&\Pi(\eta_{+},\eta_{+}^{\prime},\rho_{p})\equiv \frac{\sigma^2(\eta_{+},\rho_{p}) -\sigma^2(\eta_{+}^{\prime},\rho_{p})}{\sigma^2(\eta_{+}^{\prime},\rho_{p})},\label{eq:Pi1}\\
&&\Omega(\eta_{+},\eta_{+}^{\prime})\equiv \frac{X_{2}(\varepsilon(\eta_{+}))-X_{2}(\varepsilon(\eta_{+}^{\prime}))}
{X_{2}(\varepsilon(\eta_{+}^{\prime}))},\label{eq:Omega}
\end{eqnarray}
and then obtain the change rate associated with $\sigma^2$ and
$X_{2}(\varepsilon)$ as,
\begin{eqnarray} \label{eq:kappa}
\kappa(\eta_{+},\eta_{+}^{\prime},\rho_{p}) =
\Pi(\eta_{+},\eta_{+}^{\prime},\rho_{p}) / \Omega(\eta_{+},\eta_{+}^{\prime}),
\end{eqnarray}
where $\Omega(\eta_{+},\eta_{+}^{\prime})$ is independent of $\rho_{p}$.
In the limit $(\eta_{+}-\eta_{+}^{\prime})\rightarrow 0$, the rate of
change $\kappa(\eta_{+},\eta_{+}^{\prime},\rho_{p})$ becomes
\begin{eqnarray} \label{eq:kappa2}
\kappa(\eta_{+}^{\prime},\rho_{p})=
\frac{\partial\ln[\sigma^2(\eta_{+}^{\prime},\rho_{p})]}
{\partial\eta_{+}^{\prime}}/\frac{d \ln [X_{2}
(\varepsilon(\eta_{+}^{\prime}))]}{d \eta_{+}^{\prime}}.
\end{eqnarray}

\begin{figure}
\centering
\includegraphics[width=\linewidth]{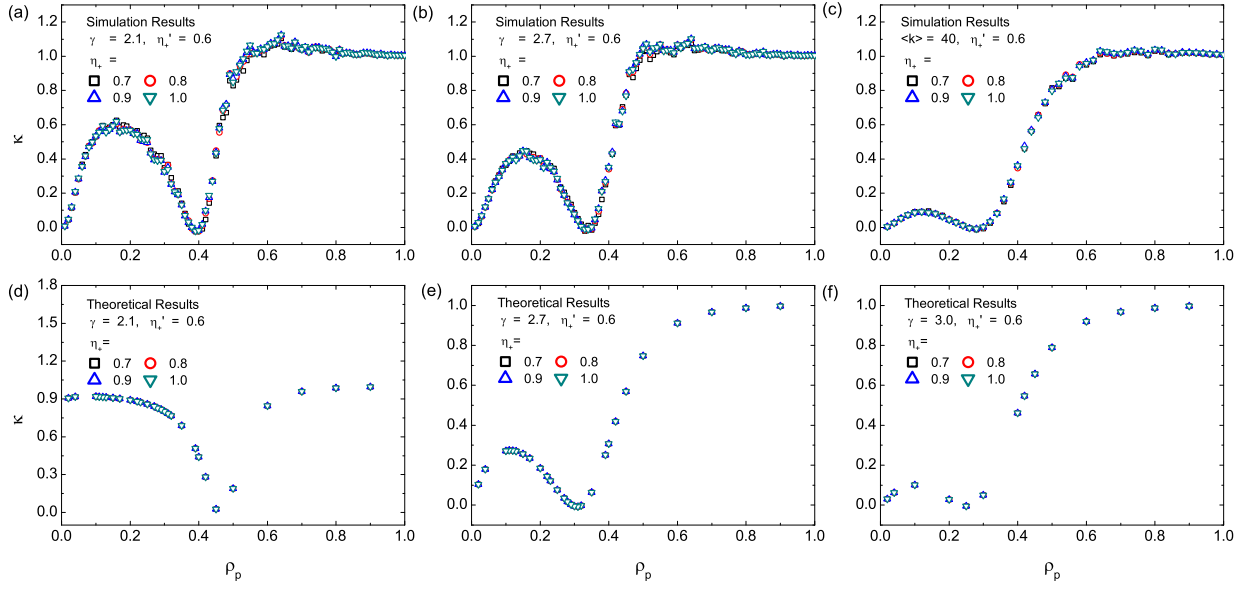}
\caption{\textbf{Collapse of $\kappa$ for different pinning patterns.}
(a-c) Simulation results of $\kappa$ from scale-free networks for
$\gamma=2.1$, $2.7$, and $3.0$, which correspond to the results of
$\sigma^2$ in Figs.~\ref{fig:2}(a),~\ref{fig:2}(c), and~\ref{fig:1}(d),
respectively. (d-f) Theoretical results of $\kappa$ from
Eq.~(\ref{eq:kappa}) for the cases shown in Figs.~\ref{fig:5}(a),
~\ref{fig:5}(c), and~\ref{fig:5}(d), respectively. The reference
pinning pattern indicator is $\eta'=0.6$.}
\label{fig:7}
\end{figure}

Figure~\ref{fig:7} shows $\kappa(\eta_{+}^{\prime},\rho_{p})$
as a function of $\rho_{p}$ for scale-free networks, where the value of
the reference pinning pattern indicator is $\eta_{+}^{\prime}=0.6$. To
obtain the values of $\kappa$, we first calculate $\Omega$
by substituting the values of $\eta_{+}$, $\eta_{+}'$ and the elements of
$\mathbb{F}$ into Eqs.~(\ref{eq:lambda}) and (\ref{eq:Omega}). We then
obtain $\Pi$ by substituting the values of $\sigma^{2}$ into
Eq.~(\ref{eq:Pi1}), with $\sigma^{2}$ either from simulation as in
Figs.~\ref{fig:1} and~\ref{fig:2} or from theoretical analysis as in
Fig.~\ref{fig:5}. We see that the $\kappa$ values from simulation
results of $\sigma^{2}$ [Figs.~\ref{fig:7}(a-c) marked by ``Simulation
Results''] and theoretical prediction of $\sigma^{2}$
[Figs.~\ref{fig:7}(d-f) marked by ``Theoretical Results''] show the
behavior in which the curves of $\kappa$ for different values of
$\eta_{+}$ collapse into a single one.
This indicates that $\kappa$ depends solely on the pinning fraction
$\rho_{p}$; it is independent of the value of the pinning pattern
indicator $\eta_{+}$. Extensive simulations and analysis of scale-free
networks with different average degree $\langle k\rangle$ or different
degree exponent $\gamma$ verify the generality of the collapsing behavior.

From Eq.~(\ref{eq:kappa2}), we see that the variance
$\sigma^2(\eta_{+},\rho_{p})$ and the quantity $\kappa$ are closely
related. For example, a smaller value of $\kappa$ indicates that
$X_{2}(\varepsilon)$ contributes more to the variance of
$\sigma^2(\eta_{+},\rho_{p})$ as $\eta_{+}$ is changed, and vice versa.
In Fig.~\ref{fig:7}, $\kappa=0$ corresponds to the intersecting points
of the curves of $\sigma^2$ with different values of $\eta_{+}$ shown
in Figs.~\ref{fig:1},~\ref{fig:2}, and~\ref{fig:5}. It can also be verified
analytically that, the minimal point with $\partial\kappa/\partial\rho_{p}=0$
coincides with the optimal pinning fraction $\rho_{p}^*$ at which
$\sigma^2$ is minimized, which is supported by simulation results in
Figs.~{\ref{fig:7}}, \ref{fig:1}, \ref{fig:2}, and \ref{fig:5}.

\begin{figure}
\centering
\includegraphics[width=\linewidth]{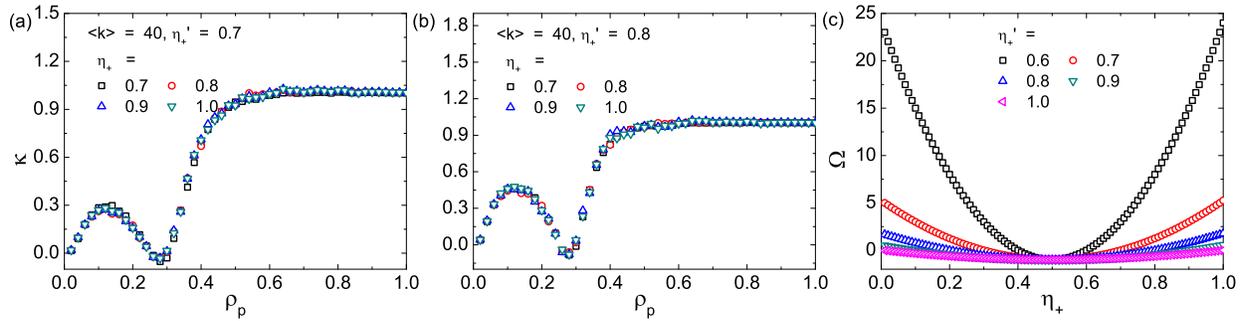}
\caption{\textbf{Two separated functions $\kappa$ and $\Omega$ in
Eq.~(\ref{eq:Pi}).} (a,b) Collapse of $\kappa(\rho_{p})$ for various
$\eta_{+}$ values, where the reference value is $\eta_{+}'=0.7$ in (a)
and $0.8$ in (b). The values of $\kappa$ are predicted from
Eq.~(\ref{eq:kappa}) for a scale-free network with $\gamma=3$ and
$\langle k \rangle=40$. (c) The function $\Omega(\eta_{+})$ for
$\eta_{+}'=0.6$, $0.7$, $0.8$, $0.9$, $1.0$, and $F_{+|-}=F_{-|+}=1$.}
\label{fig:8}
\end{figure}

\paragraph*{Variance in the form of separated variables.}
From Eq.~(\ref{eq:kappa}), for a given value of the reference pinning
pattern indicator $\eta_{+}^{\prime}$, we can obtain an expression of
$\Pi$ in the form of separated variables as
\begin{eqnarray} \label{eq:Pi}
\Pi(\eta_{+},\rho_{p})=\kappa(\rho_{p})\cdot \Omega(\eta_{+}),
\end{eqnarray}
where $\kappa(\rho_{p})$ is independent of the change in $\eta_{+}$, and
$\Omega(\eta_{+})$ is independent of $\rho_{p}$. The consequence of
Eq.~(\ref{eq:Pi}) is remarkable, since it defines in the parameter space
$(\eta_{+},\rho_{p})$ a function $\Pi$ in the form of separated variables
which, as compared with the original quantity $\sigma^2$, not only
simplifies the description but also gives a more intuitive picture of
the system behavior. Specifically, for the MG dynamics, the influences
of various factors on the variance $\sigma^2$ or $\Pi(\eta_{+},\rho_{p})$
can be classified into two parts: (I) the function $\kappa$ that reflects
the effects of the pinning fraction $\rho_p$ and the network structure
among agents (in terms of the degree distribution $P(k)$, the average degree
$\langle k\rangle$, and the scaling exponent $\gamma$), and (II) the
function $\Omega$ that characterizes the impact of the pinning pattern
indicator $\eta_{+}$ and the response of agents to resource capacities
$C_{+}$ and $C_{-}$ through $\mathbb{F}$. Figures~\ref{fig:8}(a)
and ~\ref{fig:8}(b) show the values of $\kappa$ as a function of
$\rho_{p}$ for $\eta_{+}^{\prime}=0.7$ and $0.8$, respectively, whereas
Fig.~\ref{fig:8}(c) shows $\Omega(\eta_{+})$ for several values of
$\eta_{+}^{\prime}$. From Eqs.~(\ref{eq:lambda}) and (\ref{eq:Omega}), we
see that $\Omega$ is a quadratic function of $\eta_{+}$ with the
symmetry axis at $\eta_{+}=F_{+|-}/(F_{+|-}+F_{-|+})$, which depends on the
setting of response function $\mathbb{F}$. The second derivative of the
function depends on $\eta_{+}'$.

From the definition in Eq.~(\ref{eq:Pi1}), the variance of the system
for arbitrary values of $\rho_{p}$ and $\eta_{+}$ can be obtained as
\begin{eqnarray} \label{eq:sigma2}
\sigma^2(\eta_{+},\rho_{p})=[\kappa(\rho_{p})\cdot\Omega(\eta_{+})+1]
\cdot\sigma^{2}(\eta_{+}^{\prime},\rho_{p}),
\end{eqnarray}
where $\eta_{+}^{\prime}$ specifies the reference pinning pattern.
Once we have the two respective $\sigma^2(\eta_{+},\rho_{p})$ curves
for the two specific pinning patterns as specified by $\eta_+^{\prime}$
and $\eta_{+}^{\prime\prime}$, $\sigma^2$ in the whole
parameter space $(\eta_{+},\rho_{p})$ can be calculated accordingly.
In particular, the quantities $\sigma^2(\eta_{+}^{\prime},\rho_{p})$
and $\sigma^2(\eta_{+}^{\prime\prime},\rho_{p})$ serve as a
\emph{holographic} representation of the dynamical behavior of the
system in the whole parameter space. In particular, one can first
obtain $\Omega(\eta_{+})$ from Eqs.~(\ref{eq:deltap}) and (\ref{eq:Omega}),
and then calculate
$\kappa(\rho_p)=\Pi(\eta_{+}^{\prime\prime},\rho_p)
/{\Omega(\eta_{+}^{\prime\prime})}$, and finally obtain the value of
$\sigma^2(\eta_{+},\rho_{p})$ by substituting $\Omega(\eta_{+})$ and
$\kappa(\rho_p)$ into Eq.~(\ref{eq:sigma2}). \\
\\ \noindent
{\large\textbf{Discussions}} \\
\\ \noindent
The phenomenon of herding is ubiquitous in social and economical systems.
Especially, in systems that involve and/or rely on fair resource allocation,
the emergence of herding behavior is undesirable, as in such a state a
vast majority of the individuals in the system share only a few resources,
a precursor of system collapse at a global scale. A generic
manifestation of herding behavior is relatively large fluctuations in
the dynamical variables of the system such as the numbers of individuals
sharing certain resources. It is thus desirable to develop effective control
strategies to suppress herding. An existing and powerful mathematical
framework to model and understand the herding behavior is minority games.
Investigating control of herding in the MG framework may provide useful
insights into developing more realistic control strategies for real-world
systems.

Built upon our previous works in MG systems~\cite{ZWZYL:2005,ZHDHL:2013},
in this paper we articulate, test, and analyze a general pinning strategy
to control herding behavior in MG systems. A striking finding is the
universal existence of an optimal pinning fraction that minimizes the
fluctuations in the system, regardless of system details such as
the degree of homogeneity of the resource capacities, topology and
structures of the underlying network, and different patterns of pinning.
This means that, generally, the efficiency of the system can be optimized
for some relatively small pinning fraction. Employing the
mean-field approach, we develop a detailed theory to understand and
predict the dynamics of the MG system subject to pinning control, for
various network topologies and pinning schemes. The key observation
underlying our theory is the two factors contributing to the system
fluctuations: intrinsic dynamical
fluctuations and systematic deviation of agents' expected attendance
from resource capacity. The theoretically predicted fluctuations
(quantified by the system variance) agree with those from direct simulation.
In particular, in the large degree limit, for a variety of combinations
of the network and pinning parameters, the numerical results approach
those predicted from our mean field theory. Our theory also correctly
predicts the optimal pinning fraction for various system and control
settings.

In real world systems in which resource allocation is an important
component, resource capacities and agent interactions can be diverse and
time dependent. To develop MG model to understand the effects of diversity
and time dependence on herding dynamics, and to exploit the understanding
to develop optimal control strategies to suppress or eliminate herding are open
issues at the present. Furthermore, implementation of pinning control in
real systems may be associated with incentive policies that provide
compensations or rewards to the pinned agents. How to minimize the
optimal pinning fraction then becomes an interesting issue. Our results
provide insights into these issues, and represent a step toward the
goal of designing highly stable and efficient resource allocation
systems in modern society and economy.



\section*{Acknowledgement}

We thank Prof. L. Huang for helpful discussions. This work was supported by ARO under Grant W911NF-14-1-0504, and the NSF of China under Grants
Nos. 11575072, 11135001 and 11275003.

\section*{Author contributions}

ZGH and YCL devised the research project. JQZ, ZGH, and RS performed numerical simulations. JQZ, ZGH, YCL, and ZXW analyzed the results and wrote the paper.

\section*{Additional information}

{\bf Competing financial interests}:
The authors declare no competing financial interests.


\clearpage


\end{document}